\RequirePackage{tikz}

\documentclass[aip,jcp,reprint]{revtex4-1}

\usepackage{tikz}

\usepackage{epsfig}
\usepackage{color}
\usepackage{amssymb}

\usepackage{hyperref}
\usepackage{amsmath}
\usepackage{array}
\usepackage{bm}

\definecolor{red}{rgb}{1,0,0}
\definecolor{blue}{rgb}{0,0,1}
\definecolor{black}{rgb}{0,0,0}

\newcommand{\be}{\begin{equation}}
\newcommand{\ee}{\end{equation}}
\newcommand{\ba}{\begin{eqnarray}}
\newcommand{\ea}{\end{eqnarray}}

\newcommand{\eq}[1]{\begin{align}#1\end{align}}
\newcommand{\eqs}[1]{\begin{align*}#1\end{align*}}

\def\half{\mbox{$\frac{1}{2}$}}

\def\tr{\mbox{tr}}

\newcommand{\p}{\partial}

\newcommand{\dz}{\delta z}
\newcommand{\dphi}{\delta \phi}
\newcommand{\PP}{\mathbb{P}}
\newcommand{\Gb}{\hat{G}}
\def\lh{\langle h \rangle}
\def\rb{\vec{r}}
\def\nb{\vec{n}}
\def\rb{\vec{r}}
\def\qb{\vec{q}}
\def\eff{e\!f\!f}

\newcommand{\fbar}{f_0}

\newcommand{\OO}{\mathcal{O}}

\def\kpa{k^\parallel}
\def\kpee{{\tilde k^\perp}}
\def\kpe{k^\perp}
\def\Gpa{G^\parallel}
\def\Gpe{G^\perp}
\def\curlyM{{\cal M}}
\def\delb{\hat{\delta}}

\usepackage{pgfplots}

\begin{document}

\title{Theory of the Jamming Transition at Finite Temperature}

\author{E. DeGiuli}
\affiliation{New York University, Center for Soft Matter Research, 4 Washington Place, New York, NY, 10003, USA }
\author{E. Lerner}
\affiliation{New York University, Center for Soft Matter Research, 4 Washington Place, New York, NY, 10003, USA }
\affiliation{Institute for Theoretical Physics, Institute of Physics, University of Amsterdam, Science Park 904, 1098 XH Amsterdam, The Netherlands }
\author{M. Wyart}
\affiliation{New York University, Center for Soft Matter Research, 4 Washington Place, New York, NY, 10003, USA }


\date{\today}

\begin{abstract}

A theory for the microscopic structure and the vibrational properties of soft sphere glass at finite temperature is presented. With an effective potential, derived here, the phase diagram and vibrational properties are worked out around the Maxwell critical point at zero temperature $T$ and pressure $p$. Variational arguments and effective medium theory identically predict a non-trivial temperature scale $T^*\sim p^{(2-a)/(1-a)}$ with $a \approx 0.17$ such that low-energy vibrational properties are hard-sphere like for $T \gtrsim T^*$, and zero-temperature soft-sphere like otherwise. 
However, due to crossovers in the equation of state relating $T$, $p$, and the packing fraction $\phi$, these two regimes lead to four regions where scaling behaviors differ when expressed in terms of $T$ and $\phi$. Scaling predictions are presented for the mean-squared displacement, characteristic frequency, shear modulus, and characteristic elastic length in all regions of the phase diagram.

\end{abstract}
\maketitle

 A wide range of amorphous materials including granular materials, foams, molecular glasses and colloids exhibit a  transition from liquid-like to solid-like behavior. In the solid phase, these materials display anomalous elastic properties. In particular, amorphous solids universally present an excess of vibrational modes over the Debye model (that treats vibrational modes as plane waves) a phenomenon referred to as the `boson peak' \cite{Phillips81}. The boson peak affects various properties of the solid, including heat transport \cite{Baldi10,Baldi11,Kittel49,Vitelli10,Xu09} as well as the spatial heterogeneity of the elastic  response \cite{Tanguy02,Leonforte06,Silbert05,Lerner14}. Its presence  indicates that glasses lie close to an elastic instability, suggesting a connection with the glass transition  where rigidity emerges \cite{Grigera02,Parisi03}. For these reasons this phenomenon is intensely studied \cite{Chumakov15}.

It was proposed that in a variety of glasses \cite{Wyart05b,DeGiuli14}, the boson peak is controlled by some aspects of the short-range microscopic structure, in particular a measure of the particle connectedness \cite{Trachenko04,Wyart05,Xu07} as well as the characteristic force with which nearby particles interact  \cite{Alexander98,Wyart05a}, both well-known to affect the stability of engineering structures \cite{Maxwell64}. Detailed predictions on transport and sound dispersion  can be made in this framework \cite{DeGiuli14}, which have been recently supported by experiments \cite{Buchenau14}.
In this approach, particles interacting via a purely repulsive potential at zero temperature \cite{Hecke10,Liu10,Ohern03} are predicted to display singular vibrational properties: the spectrum has a characteristic frequency $\omega^*$  satisfying $\omega^* \sim z-z_c$ near the jamming transition \cite{Wyart05}, as confirmed numerically \cite{Silbert05}. Here $z$ is coordination, measured by the average number of contacts per particle,  and  $z_c=2d$ is the minimal coordination for mechanical stability  in dimension $d$, the so-called Maxwell threshold.  The boson peak frequency $\omega_{BP}$, however, is essentially zero, with $\omega_{BP}\ll \omega^*$ indicating that the system is {\it marginally stable} \cite{Wyart05a} and implying that $z-z_c\sim \sqrt{\phi-\phi_c}$, where $\phi_c$ is the jamming packing fraction where pressure vanishes (that can depend on the configuration considered). 

\begin{center}
\begin{table*}[!t]
\newcolumntype{B}{ >{\centering\arraybackslash} p{1.2cm} }
\newcolumntype{C}{ >{\centering\arraybackslash} p{1.5cm} }
\newcolumntype{D}{ >{\centering\arraybackslash} p{2.5cm} }
\begin{tabular}{ p{3cm} | D C | C B C C C C}
 Regime & $T$ & $\dphi$ &  $p$  &  $\lh$  & $\omega^*$ & $\langle \delta R^2 \rangle$ & $\mu$ & $\ell_s(\omega^*)$ \\
\hline
Hard-sphere (HS) & $T \ll \dphi^2$ & $\dphi < 0$ & $T/|\dphi|$ & $|\dphi|$ & $(T/|\dphi|)^{1/2}$ & $|\dphi|^{\kappa}$ & $T |\dphi|^{-\kappa}$ & $|\dphi|^{(a-1)/4}$ \\
Soft-zero-T (S0) & $T^* \ll T \ll \dphi^2$ & $\dphi > 0$ & $\dphi$ & $-\dphi$ & $\dphi^{1/2}$ & $T |\dphi|^{-1/2}$ & $|\dphi|^{1/2}$ & $\dphi^{-1/4}$\\
Soft-entropic (SE) \quad & $T \ll T^*$ & $\dphi > 0$ & $\dphi$ & $-\dphi$ & $\dphi^{1/2}$ & $(T/|\dphi|)^{\kappa}$ & $T^{1-\kappa} |\dphi|^{\kappa}$ & $(T/\dphi)^{(a-1)/4}$ \\
Anharmonic (AH) & $T \gg \dphi^2$ & & $\sqrt{T}$ & $\sqrt{T}$ & $T^{1/4}$ & $T^{\kappa/2}$ & $T^{1-\kappa/2}$ & $T^{(a-1)/8}$ \\
\end{tabular}
\caption{\label{tab1} Regimes of the soft-sphere glass. Shown is the scaling behavior of pressure $p$, typical time-averaged gap $\langle h \rangle$, characteristic frequency $\omega^*$, mean-squared displacement $\langle \delta R^2 \rangle$, intra-state shear modulus $\mu$, and correlation length of the normal modes $\ell_s(\omega^*)$ versus temperature $T$ and volume fraction deviation $\dphi = \phi - \phi_c$. As discussed in the main text, the soft regime contains a transition at a nontrivial temperature scale $T^* \sim \dphi^{(5+3\theta_e)/(2+2\theta_e)}\sim \dphi^{2.2}$, invisible in the effective potential and equation of state (and therefore $p$ and $\lh$), but strongly affecting vibrational properties. The exponents $\kappa = (4+2\theta_e)/(3+\theta_e)\approx 1.41$ and $a=(1-\theta_e)/(3+\theta_e) \approx 0.17$, where $\theta_e \approx 0.42$ characterizes the force distribution of ``extended" contacts at jamming \cite{Lerner13a,Charbonneau14a,Charbonneau14,Charbonneau14c}. } 
\end{table*}
\end{center}

Since in practice interaction potentials are always anharmonic, one may question if these results survive at finite temperature \cite{Schreck11,Henkes12}. Hard spheres, arguably the simplest glass former, are a particularly challenging case for which the potential is discontinuous, implying that the Hessian of the energy (whose eigenvectors correspond to the normal modes) is not well-defined.  
However, it was argued that a coarse-grained free energy can be computed in that case \cite{Brito06,Brito09,DeGiuli14b} if one can find times $\tau$ such that $\tau_c \ll \tau \ll \tau_\alpha$, where $\tau_c$ is a microscopic collision time, and $\tau_\alpha$ is the relaxation time of the structure (this condition is always achieved in the glass phase).  This free energy captures the volume of phase space around a given meta-stable state (or ``vibrational entropy"), and can always be expressed in terms of the mean particle position within a meta-stable state. This approach leads to an effective interaction $V_{\eff}$ between particles, which is in general multi-body. Near the jamming point, however, this interaction is simply two-body (as recalled below), and only occurs between particles interacting  (i.e. colliding) on the intermediary time $\tau$, leading to the definition of a contact network and allowing to extend the notion of coordination to hard spheres. 
The Hessian of the coarse-grained free energy describes the linear response, in particular how small external forces affect the mean particle positions.  Its eigenvectors define normal modes whose thermal fluctuations are inversely proportional to their associated eigenvalue, as confirmed numerically \cite{Brito07}.

Theoretical arguments using such an effective potential were used to predict scaling properties of the spectrum of the Hessian, as well as spatial properties of the eigenvectors  \cite{Brito09,DeGiuli14b}. One finds that the normal modes slightly differ from the soft sphere case: in both cases  modes are extended and heterogeneous, but for hard spheres soft modes tend to distort  a small amount of  bonds (which are weak), whereas for soft spheres modes tend to distort all the bonds in the system. We thus coin hard-sphere soft modes as {\it sparse}, whereas for soft spheres we keep the previously used terminology of {\it anomalous modes}. This distinction leads to a small but detectable \cite{DeGiuli14b} difference in vibrational properties. For the coordination, it was predicted that $z-z_c\sim (\phi_c-\phi)^{0.41}$. Other predictions, such as for the mean-squared displacement, were found to agree with recent replica calculations  in infinite dimensions \cite{Charbonneau14a,Charbonneau14}, supporting the validity of the two approaches. 

\begin{figure}[b!] 
\includegraphics[width=0.48\textwidth,clip]{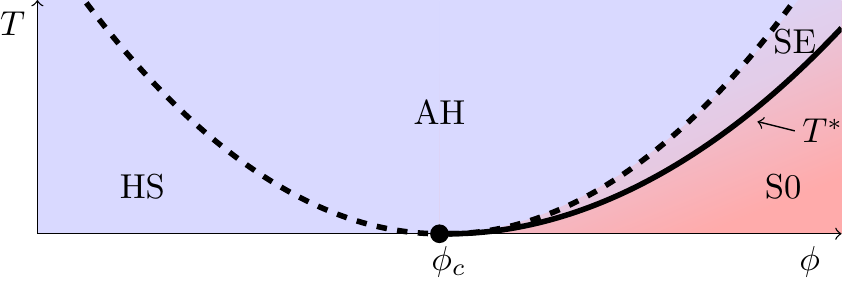}
%
%
\caption{\label{fig:phasediagram} Phase diagram for marginally-stable harmonic soft sphere systems near the Maxwell point at $(T,\phi)=(0,\phi_c)$. The dashed line indicates the scale $T \sim |\dphi|^2$, while the solid line indicates $T^* \sim \dphi^{(5+3\theta_e)/(2+2\theta_e)}$. In regimes Hard Spheres (HS), Anharmonic (AH), and Soft Entropic (SE), the softest modes are `sparse,' as in hard sphere systems, but with different scalings for the pressure: $p \sim T/|\dphi|$ (HS), $p \sim \sqrt{T}$ (AH), and $p \sim \dphi$ (SE). In regime S0, the softest modes are `anomalous,' like $T=0$ soft solids. Note that the lines are schematic. }
\end{figure}

The asymmetry between soft and hard spheres apparent in the scaling of coordination is surprising,  since one can go continuously from one case to the other by considering soft spheres, and by allowing to vary both $\phi$ and $T$ \cite{Schreck11,Otsuki12,Henkes12,Ikeda12,Wang13,Ikeda13,Olsson13,Bertrand14}. The limit $\phi>\phi_c$, $T\rightarrow 0$ corresponds to soft spheres whereas hard spheres are recovered in the limit $\phi<\phi_c$, $T\rightarrow 0$. Previous theoretical attempts \cite{Berthier11a,Jacquin11,Ikeda13} to describe vibrational properties in the $(\phi,T)$ plane did not include this necessary asymmetry. 
Here we extend the effective potential approach to the case of thermal soft spheres, and use our real-space description of vibrational modes to describe the phase diagram entirely.
Our main results are summarized in the phase diagram of Fig.\ref{fig:phasediagram}, and Table \ref{tab1}. We find that low-energy vibrational properties display a single crossover scale, $T^*$, when written in terms of pressure $p$ and temperature $T$; this transition is indicated by the shaded color change from red to blue in Fig.\ref{fig:phasediagram}. However, if pressure is eliminated in favor of volume fraction $\phi$, this introduces two more transitions, reflecting crossovers in the equation of state. In sum, this gives four scaling regimes in the phase diagram.

%

The definition of a contact network and effective potential we provide complements the approach of Ref. \onlinecite{Henkes12}, who discussed anharmonicity in amorphous solids from an experimentalist's point of view. The experimental question is: given the displacement covariance matrix for an anharmonic system, what can one learn about the vibrational modes \cite{Ghosh10,Chen10}?  The authors of Ref. \onlinecite{Henkes12} defined the `shadow system,' which is the Hamiltonian system of harmonic springs whose stiffnesses are related to displacements by the usual equilibrium formula. The obtained effective stiffnesses imply an interaction between particles on the averaging time-scale used to obtain displacement correlations. As we will see, we find a remarkable similarity between our effective potential and the inferred potential obtained from particle displacements in the numerical simulations of Ref. \onlinecite{Henkes12}.



\section{ Effective Potential }
\subsection{ Derivation }
We consider a system of $N$ soft spheres at constant pressure $p$ that is fluctuating around a  metastable state. { Thermal averages taken around this state are written by $\langle \cdot \rangle$, while averages over the contacts will be denoted by $\overline{\;\cdot\;}$. } We assume that the coordination of the network of interacting particles is $z_c =2d$, i.e the system is isostatic, as is the case at the jamming transition $p, T \to 0$. The computation of the vibrational free energy in such a meta-stable state is  detailed in Appendix A, and here we sketch the argument. Isostaticity implies that the number of degrees of freedom in the particle displacements $\{\delta \rb_i \}$, $dN$, is precisely equal to the number of contacts, $N_C=zN/2$. Thus the partition function of the metastable state, originally an integral over the displacements $\{\delta \rb_i\}$, can be written as an integral over the gaps $\{h_{\alpha}\}$ between contacting particles (with $h_{\alpha}<0$ for overlap). Assuming that all $|\delta \rb_i | \ll 1$, the map from $\{\delta \rb_i \}$ to $\{h_{\alpha}\}$ is linear. 

To compress or dilate the system from a volume $V_c$ (where all particles are just touching) to $V$ requires work $W=-p(V-V_c)$. Stability implies that the time-averaged forces $f_\alpha$ between particles satisfy force balance; this can be written 
 $W = -\sum_\alpha f_\alpha h_\alpha$, where the sum is over all contacts $\alpha$ of the metastable state, and we have used the virtual work principle \cite{Roux00}. Since $W$ and the elastic energy $U$ are sums over contributions from different contacts (we consider only pair potentials), this leads to a single-gap partition function \cite{Brito06,Brito09}
\eq{ \label{Z}
Z(\beta,f) = e^{-\beta G} = \int dh \; e^{-\beta U(h)} e^{-\beta f h},
}
where the force $f$ fixes the time-averaged gap $\lh$ through $\lh = \p G/\p f$, and $\beta=1/(k_B T)$. The effective potential $V_{\eff}$ is obtained by a Legendre transform $V_{\eff}(\lh) = G(f)-f \lh$, from which it follows that $f = -\p V_{\eff}(\lh)/\p \lh$. We consider a finite-range harmonic potential $U(h)=\half\epsilon |h/\sigma|^2$ when $-\half \sigma < h < 0$, and 0 otherwise. We now take units in which $\epsilon=\sigma=k_B=1$. As described in Appendix A, it is simple to analytically extract the limiting behavior: when $f \ll \sqrt{T}$, $\lh \approx T/f$, whereas when $f \gg \sqrt{T}$, $\lh \approx -f$. These correctly reduce to the effective hard-sphere potential from Refs. \onlinecite{Brito06,Brito09} in the small-force limit, and the original harmonic potential in the large-force limit, respectively, and moreover they indicate the cross-over scales. In the intermediate regime, $|\lh| \lesssim \sqrt{T}$, there is a smooth cross-over between these behaviors. 

\begin{figure}[t] 
\includegraphics[width=0.48\textwidth,clip]{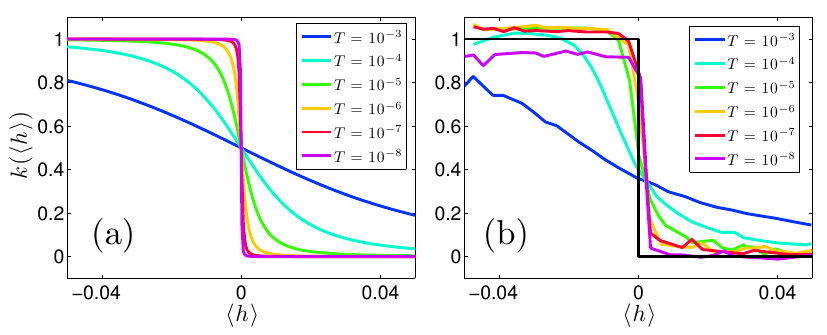}
\caption{\label{fig:keff} Effective stiffness at indicated temperatures (a) as predicted from the simple form Eq.\eqref{keff}, and (b) as computed from the displacement covariance matrix in numerical simulations \cite{Henkes12} at $\phi-\phi_c=0.02$. }
\end{figure}

Since our goal is to obtain scaling behavior, it is convenient to have a simple form that incorporates the salient features of $V_{\eff}$, without adding extra structure. The simplest is the one satisfying $\lh = -f + T/f$, which leads to an effective force law
\eq{ \label{feff}
f(\lh) = -\half \lh + \half \sqrt{\lh^2+4T},
}
from which we can compute the effective potential $V_{\eff}(\lh) = -\half\lh f(\lh) - T \log[\lh+\sqrt{\lh^2+4T}]$ and stiffness
\eq{ \label{keff}
k(\lh) \equiv & -\frac{df}{d\lh} = \frac{1}{2} - \frac{\lh}{2\sqrt{\lh^2+4T}}
}
The effective stiffness is plotted in Fig.~\ref{fig:keff}a, showing an entropic smoothing as $T$ is increased from zero. In the introduction, it was argued that the effective potential can be measured by displacement covariances using the usual harmonic formula \cite{Henkes12}. To test this claim explicitly, in Fig. \ref{fig:keff}b we reproduce the effective stiffness $k_{\eff}$ extracted from displacement covariances in Ref. \onlinecite{Henkes12} from numerical simulations of a thermal soft-sphere system in $d=2$. A very good qualitative agreement is obtained, with no fitting parameters. Moreover, in Ref. \onlinecite{Henkes12} it was observed that by varying $\dphi=\phi-\phi_c$ from $\approx -0.02$ to $+0.05$, the effective stiffness function $k_{\eff}(\lh)$ remained constant, and the different systems merely sampled different ranges of $\lh$; this feature is reproduced by Eq.\eqref{keff}, which has no explicit dependence on $\phi$. This observation supports that corrections to the effective potential away from isostaticity are small in this range of $\dphi$, as was checked previously for hard spheres \cite{Brito06}.

\subsection{Force Distribution}

{ The disordered geometry of the metastable state is characterized by the statistics of time-averaged particle positions, which determines in turn the distribution of forces between particles. }
In Ref. \onlinecite{Lerner12}  the force distribution in packings of frictionless particles at jamming was found to be singular:
 \eq{ \label{pdff}
\PP(f) \propto f^{\theta_\ell} e^{-f/\fbar}}
as confirmed by several numerical studies \cite{Charbonneau12,Lerner13a,DeGiuli14b,Charbonneau14c}. Note that the exponential tail was observed earlier in Ref. \onlinecite{Donev05a}, but is not crucial; a Gaussian cut-off would not change our scaling predictions below. The singularity at small $f$ was shown to be necessary for the stability of packings toward rewiring of their contact network \cite{Wyart12,Lerner13a}. More precisely, contacts need to be classified in two groups. {\it Local contacts} have very little mechanical coupling with their surroundings, and their density follows Eq.(\ref{pdff}). {\it Extended contacts} are coupled to their surroundings, and their density follows:
\be
\label{ex}
\PP_e(f)\sim f^{\theta_e} e^{-f/\fbar}.
\ee
The latter are less numerous at low force, i.e. $\theta_e>\theta_\ell$, but more important for our purpose \footnote[1]{Cutting extended contacts as in the variational argument leads to extended floppy modes with a much lower variational energy than by cutting localized contacts, which leads to localized floppy modes.}. Marginal stability of packings implies \cite{Lerner13a,Muller14}:
\be
\label{mar}
\theta_\ell=\frac{\theta_e}{2+\theta_e}
\ee
in agreement with numerics indicating $\theta_\ell\approx 0.17$ and $\theta_e\approx 0.44$. 
Replica calculations\cite{Kurchan12,Kurchan13,Charbonneau14a,Charbonneau14} in $d=\infty$ do not predict $\theta_\ell$ but yield $\theta_e=0.42311..$ \cite{Charbonneau14}, consistent with the value extracted from numerical observations in finite dimensions.


%
%

In what follows, we assume that the distribution of Eq.(\ref{ex}) holds true in the vicinity of the jamming transition. Note that away from the Maxwell point, $\PP_e(f)$ develops a plateau at small force \cite{Charbonneau12,Charbonneau14}.  For simplicity, we neglect this plateau in our arguments, since as shown previously in the hard sphere regime \cite{DeGiuli14b} it is not expected to affect our results.

\begin{figure}[!t] 
\includegraphics[width=0.50\textwidth,clip]{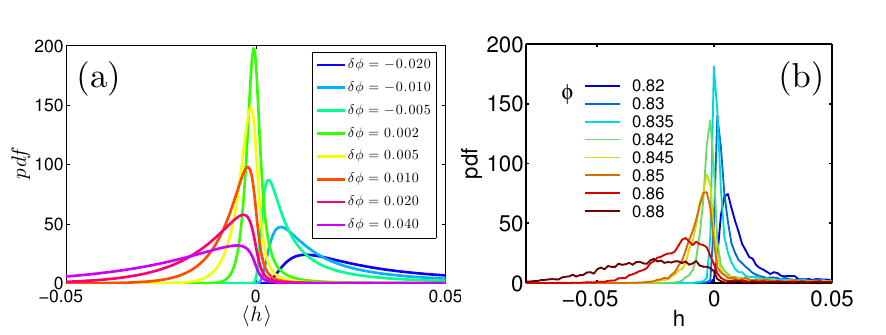}
\caption{\label{fig:pdfh} Time-averaged gap distribution at $T=10^{-6}$ and indicated $\delta\phi$ (a) as predicted from Eq.\eqref{feff}, Eq.\eqref{pdff}, and Eq.\eqref{eqstate}, and (b) as computed from the displacement covariance matrix in numerical simulations \cite{Henkes12}. }
\end{figure}



Assuming Eq.\eqref{pdff} with\cite{Lerner12,Lerner13a,Charbonneau14c} $\theta_\ell=0.175$ and the effective potential Eq.~\eqref{feff} immediately yields a prediction for the time-averaged gap distribution, as a function of $\fbar$ and $T$. Using the equation of state derived below, this can be written as a function of $\dphi$, plotted in Fig.\ref{fig:pdfh}a at $T=10^{-6}$. As $\dphi$ is varied, the distribution has a complex evolution. Comparing this to numerical results from Ref.\onlinecite{Henkes12}, plotted in Fig.\ref{fig:pdfh}b, again a very good qualitative agreement is obtained, with no fitting parameters. Moreover, the scaling behavior of $\PP(h)$ can easily be extracted. One finds in particular that $\PP(h)$ has a maximum at $h^\dagger \sim \overline{\lh}$, where $\overline{\; \cdot \;}$ denotes the average over contacts, i.e., with respect to Eq.\eqref{pdff}. { By averaging $\lh = -f + T/f$ over contacts, the mean gap is 
\eq{ \label{hbar}
\overline{\lh} = -(1+\theta_\ell) \fbar + T/(\fbar \theta_\ell).
}
The height of the maximum of $\PP(h)$} scales as $\PP(h^\dagger) \sim (\fbar + T/\fbar)^{-1}$. 

From $\PP(h)$ one can also extract the instantaneous coordination $z_{inst}$ by $z_{inst}/z=\int_{-1/2}^0 dh \; \PP(h)$. The result is that in most of the phase diagram, $z_{inst}<z_c$: the system is instantaneously hypostatic. However, the instantaneous coordination is not the correct variable to characterize stability and vibrational properties-- this role is played by the coordination coarse-grained in time. 

\subsection{Equation of State}

The effective potential and force distribution also imply the equation of state. { The Irving-Kirkwood expression for the stress tensor\cite{Irving50} gives its contact contribution as $pV/N_{C}=\overline{(1+h)f}$, whose thermal average is $p \langle V \rangle/N_C = \overline{(1+\lh )f}$. Meanwhile, mechanical stability of the metastable state allows to write the work done in compression from $V_c$ in terms of forces and gaps. The latter expression, the virtual work principle derived in Appendix A, equates $p(V-V_c)$ to its microscopic expression $N_C \overline{ h f }$; the thermal average is $p(\langle V \rangle -V_c)/N_C = \overline{ \lh f}$. Subtracting these expressions we find $pV_c/N_C = \overline{f} = (1+\theta_\ell) f_0$, using Eq.\eqref{pdff}. Similarly, we use the effective potential $\lh = -f + T/f$ and Eq.\eqref{pdff} to evaluate $\overline{ \lh f} = \overline{-f^2+T} = -(2+\theta_\ell)(1+\theta_\ell)f_0^2 + T$. Substituting this into the virtual work principle and simplifying with $\langle V \rangle = V$ and $(V_c-V )/ V = \dphi/\phi$, valid in the thermodynamic limit, we find eventually the equation of state
\eq{ \label{eqstate}
p \frac{V_c}{(1+\theta_\ell) N_C} = \fbar = \frac{ \dphi +\sqrt{(\dphi)^2+c_1 T \phi^4 }}{2\phi^2 \phi_c^{-1} (2+\theta_\ell)},
}
where $c_1=4(2+\theta_\ell)/(\phi_c^2 (1+\theta_\ell))$. }
Note that $\overline{f} = (1+\theta_\ell)\fbar$ and in all cases $p \sim \fbar$. From Eq.\eqref{eqstate}, we see a temperature scale $T \sim |\dphi|^2$ delineating three regimes: a hard-sphere like regime (HS), a soft regime (S), and an anharmonic regime (AH) as indicated in Table 1. These are denoted `unjammed scaling,' `jammed scaling,' and `finite temperature critical' in Ref.\onlinecite{Ikeda13}, respectively. 

In fact, we now show that in the soft regime, temperature plays a role at a {\it smaller}, nontrivial  scale. The former is thus actually two regimes: soft-zero-temperature (S0), and soft-entropic (SE). Their scalings are summarized in Table \ref{tab1}. 

%
%




\begin{figure}[t!] 
\includegraphics[width=0.48\textwidth,clip]{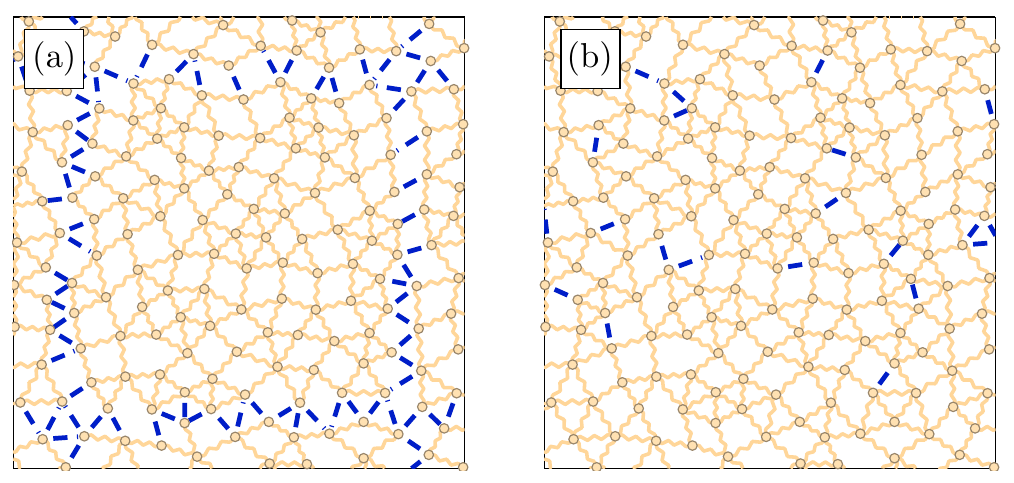}
\caption{\label{fig:cut} Illustrative diagram of cutting argument, showing cut bonds in blue (thick lines). (a) Bonds are cut around blocks of size $L\times L$, a useful procedure at small $T$; (b) When $T$ is large enough, the variational argument is improved by cutting instead the fraction $q$ of weakest extended bonds. From Ref. \onlinecite{DeGiuli14b}. }
\end{figure}

\section{Variational Arguments}
\subsection{ Density of states } The effective potential defines a harmonic shadow system whose density of vibrational states, $D(\omega)$, can be bound with a variational argument \cite{Wyart05,Wyart05b}. We follow Ref. \onlinecite{DeGiuli14b}, sketching only the salient modifications from the $T\to 0$ arguments presented there.  

The shadow system is equivalent to an elastic network with a fixed time-averaged coordination $z\geq z_c$. Our goal is to construct normalized trial displacement fields with a small elastic energy $E$ as measured by the stiffness matrix of the elastic network. It is customary to use the notation $E=\omega^2$, although our predictions, based on thermodynamics, apply both to over-damped Brownian particles, or to inertial ones. 

If $Q$ orthonormal modes per unit volume can be found with a characteristic frequency $\omega$, a variational inequality implies that\cite{Wyart05} $D(\omega) \gtrsim Q/\omega$, where here and in the remainder of this section we ignore unimportant constants. To construct trial modes, a useful ansatz near the Maxwell point is to cut a fraction $q \ll 1$ of bonds, creating a density $q-\dz/z_c$ of floppy (i.e. zero-energy) modes in the cut system. These floppy modes have motion transverse to all bonds, but motion parallel to bonds {\it only} at the cut bonds. In the original, uncut system, these modes compress or extend springs at the cut bonds, but a judicious choice of cut bonds can lead to small energy and an optimal bound on $D(\omega)$. Two ansatze have proven useful: one that creates `anomalous' modes \cite{Wyart05}, and another that creates `sparse' modes \cite{DeGiuli14b}. 
First we suppose that the springs are at their rest length. 

\begin{figure}[t!] 
\includegraphics[width=0.46\textwidth,clip]{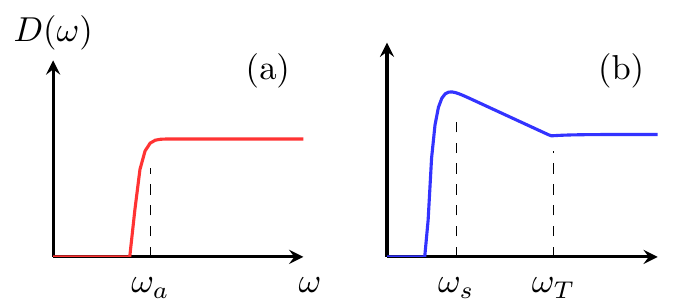}
%
%
\caption{\label{sketch} Sketch of bound on $D(\omega)$ from variational argument, (a) for $T<T^*$, and (b) for $T>T^*$, in log-log axes. }
\end{figure}

\subsection{Anomalous modes} As discussed in Ref. \onlinecite{Wyart05b}, trial modes can be created by cutting bonds along blocks of size $L\sim1/q \gg 1$ (see Fig. \ref{fig:cut}a). The induced floppy modes have large motions at the cut bonds; by modulating these modes with plane waves with nodes at the cut bonds, the resulting trial modes have a small frequency $\omega \sim q/\sqrt{{\bar k}}$. It turns out that in the regimes where anomalous modes can dominate the low-frequency spectrum, ${\bar k}\sim1$ and $\omega \sim q$. This leads to a bound:
\be
\label{ano}
D(\omega) \gtrsim \omega^0\equiv D_a(\omega),
\ee
valid above the characteristic frequency $\omega_a \sim \dz$. This is sketched in Fig.\ref{sketch}a. Since this argument is purely geometrical, it applies without change in the case of an effective potential, as considered here.

\subsection{Sparse modes} The second ansatz, discussed in Ref. \onlinecite{DeGiuli14b}, creates trial modes that only extend and compress a fraction $q$ of the weakest bonds; we call these modes `sparse'. 
 The trial modes created by cutting these contacts have a characteristic stiffness $k_c$, and characteristic force $f_c$, satisfying $q = \int_0^{f_c} \PP(f) df$. This leads to $f_c \sim p q^{1/(1+\theta_e)}$. From Eq.\eqref{feff} and Eq.\eqref{keff} we find $k_c~\sim~p^2 q^{2/(1+\theta_e)} / (T+p^2 q^{2/(1+\theta_e)})$, so that the elastic energy of the trial modes is $E \sim q k_c$. These two scales are the only ingredients needed from the effective potential of the preceding section; in particular, the equation of state does not directly affect vibrational properties. 
 
{ Although in principle both localized and extended contacts could be cut, it was checked in Ref. \onlinecite{DeGiuli14b} that only the extended contacts give a useful bound for hard spheres. We therefore use only them here, taking $\PP(f)$ from Eq.\eqref{ex}. } 
 
To obtain a useful bound from this ansatz, the characteristic stiffness must be small; in particular, we must have $T/p^2 \gg q^{2/(1+\theta_e)}$ (otherwise $k_c \sim 1$). This can be written $q \ll q_T$ with $q_T \sim (T^{1/2}/p)^{1+\theta_e}$.  
 In this case the modes have a characteristic frequency $\omega(q) \sim \sqrt{E} \sim \sqrt{q k_c} \sim p q^{1/(4b)} T^{-1/2}$, and the variational inequality gives
\eq{ \label{Domega}
D(\omega) \gtrsim \big( T/p^2 \big)^{2b} \omega^{-a}\equiv D_s(\omega),
} 
with 
\eq{
a=(1-\theta_e)/(3+\theta_e), \qquad b=(1+\theta_e)/(6+2\theta_e),
}
as obtained previously for hard spheres \cite{DeGiuli14b}. This holds for any $q \gtrsim \dz$, implying that $\omega \gtrsim \omega_s$ with
\eq{
\omega_s \sim (p/\sqrt{T}) \; \dz^{1/4b}.
}

\subsection{Optimal bound} At finite $T$, both `anomalous' and `sparse' bounds are useful, in different parts of the phase diagram, and different frequency ranges. The optimal bound on $D(\omega)$ is simply whichever is larger of Eqs.(\ref{Domega}, \ref{ano}):
\be
D(\omega)=\max(D_s(\omega),D_a(\omega))
\ee
Accordingly, the characteristic frequency of the lowest-frequency modes is $\omega^* = $ min$[\omega_a,\omega_s]$. 
The equality $\omega_a=\omega_s$ defines a characteristic temperature \eq{ T^* \sim p^2 \; \dz^{a/2b},} and therefore two cases:

Case $T<T^*$: the vibrational spectrum is dominated by anomalous modes and the density of states is flat, as occurs at zero temperature for soft particles.

Case $T>T^*$: the lowest soft modes are sparse-like. However since Eq.\eqref{Domega} is a decreasing function of $\omega$, eventually Eq.\eqref{Domega} yields $D(\omega) \gtrsim 1$, and the bounds coincide. This occurs at a frequency \footnote[2]{It can be checked that $\omega_T < \omega(q_T)$ and therefore the requirement that the characteristic stiffness is small is irrelevant.}
\eq{
\omega_T = (T/p^2)^{2b/a}.
}
So the vibrational spectrum contains both features: for $\omega_s \ll \omega \ll \omega_T$, $D(\omega)$ is hard-sphere like and decays as a power-law of frequency. However it becomes flat for $\omega_T \lesssim \omega \lesssim 1$. We include in the latter case the possibility that $\omega_T \sim 1$, in which case the flat regime is not visible. 
These two behaviors of $D(\omega)$ are sketched in Fig.\ref{sketch}. 
\begin{figure}[t] 
\includegraphics[width=0.5\textwidth,clip]{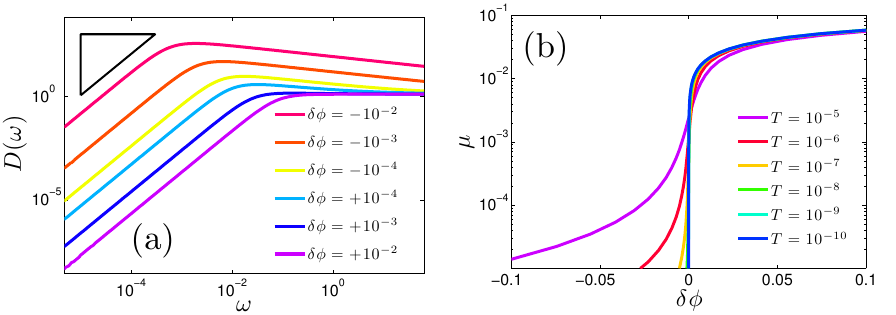}
\caption{\label{fig:domega} Effective medium theory predictions for a soft-sphere system at marginal stability. (a) Density of states $D(\omega)$ at $T=10^{-8}$ and indicated values of $\dphi$. The triangle has slope 2. (b) Shear modulus $\mu$ at indicated values of $T$.}
\end{figure}

\subsection{Coordination and marginal stability} The coordination $z$ is a natural variable to describe vibrational properties, but it is generally not experimentally accessible. Here we relate it to $T$ and $p$ by considering that the system is marginally stable, as previously  argued from dynamical considerations \cite{Brito09}, and as confirmed by replica calculations in infinite dimension \cite{Charbonneau14}. The elastic energy of a spring network under compression has two components: a positive (stabilizing) one, due to motion parallel to springs, and a negative (destabilizing) one, due to motion transverse to springs. Marginal stability implies that for the softest sparse or anomalous modes, the destabilizing transverse part of the elastic energy is of the order of the stabilizing part. 
The most unstable (anomalous or sparse) trial modes used to bound $D(\omega)$ are those at frequency $\omega^*$, with energy $ \omega^*{}^2$. For such modes, the destabilizing energy scales as the pressure\cite{Wyart05a} $-p$. We thus get $\omega^* \sim \sqrt{p}$. This sets $\dz$ and $T^*$ as 
\eq{ 
\dz \sim \begin{cases} (T/p)^{2b} \qquad & T > T^* \label{dz} \\
\sqrt{p} \qquad & T < T^* \\ \end{cases}
}
and
\eq{
T^* \sim p^{\frac{2-a}{1-a}}.
}

\subsection{Mean-squared displacement}
These results yield a bound on the particles' mean-squared displacement $\langle \delta R^2 \rangle$. \eq{
\frac{\langle \delta R^2 \rangle}{T} & = \int \frac{D(\omega)}{\omega^2}d\omega > \int_{\omega>\omega^*} \frac{D(\omega)}{\omega^2}d\omega \label{msd} \\
& \gtrsim \begin{cases} \big( T/p^2 \big)^{2b} \omega^*{}^{4b-2} \qquad & T > T^*,  \\
1/\omega^* \qquad & T < T^* \notag \end{cases} \notag \\
&\gtrsim \begin{cases} \big( T/p \big)^{1+2b} \qquad & T > T^*,  \\
T/p^{1/2} \qquad & T < T^* \notag \end{cases}
}
From effective medium theory, which makes detailed predictions for the shape of $D(\omega)$ for $\omega<\omega^*$ (see below), we predict these bounds to be saturated.

\begin{figure}[t] 
\includegraphics[width=0.5\textwidth,clip]{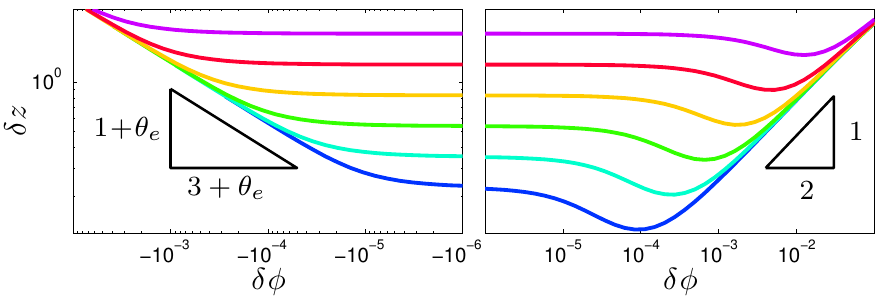}
\caption{\label{fig:dz} Effective medium theory prediction for $\dz (T,\dphi)$ at marginal stability, at various temperatures from $T=10^{-10}$ (blue) to $T=10^{-5}$ (purple). Here we take $\theta_e = 0.42$.}
\end{figure}

\section{Phase diagram $(\phi,T)$}

\subsection{ Soft regime (S0), (SE)} At $T=0$ and $\phi>\phi_c$, the system is in a `jammed scaling' regime \cite{Ikeda13} with $p \sim \dphi$,  $\dz \sim \sqrt{p}$, and a plateau of `smooth' modes above $\omega^* \sim \dz$ \cite{Wyart05, Wyart05b}. This soft-zero-temperature (S0) regime persists for $T \lesssim T^* \sim \dphi^{(2-a)/(1-a)}$, at which point the bound \eqref{Domega} first becomes applicable. For $T^* \lesssim T \lesssim \dphi^2$, the density of states has the form \eqref{Domega}, and $\dz\sim (T/p)^{2b}$, but the pressure is still dominated by contacts. This is the nontrivial regime in which {\it typical} contacts are elastic, but enough {\it weak} contacts are entropic to affect vibrational properties. We call it SE -- soft entropic. 

Collecting results, and using Eq.\eqref{msd}, we get
\eq{
\omega^* \sim \dphi^{1/2} \qquad & (S0), (SE)
}
and
\eq{ \label{msd1}
\langle \delta R^2 \rangle \gtrsim \begin{cases} T \; \dphi^{-1/2} \qquad & (S0) \\
(T/\dphi)^{\kappa} & (SE)
\end{cases},
} 
with 
\eq{
\kappa = \frac{3}{2} - \frac{a}{2} = \frac{4 + 2\theta_e}{3+\theta_e}.
}

\subsection{ Anharmonic regime (AH) }

Both above and below jamming, when the temperature is large enough, $T \gtrsim \dphi^2$, the pressure is entropic: $p \sim \sqrt{T}$. In this regime
\eq{
\omega^* \sim T^{1/4} \qquad & (AH)
}
and
\eq{ \label{msd1b}
\langle \delta R^2 \rangle \gtrsim T^{\kappa/2} \qquad & (AH)
} 

\subsection{ Hard-sphere regime (HS) }

When $\dphi < 0$, as $T\to 0$ the system behaves like hard spheres, with $p \sim T/|\dphi|$, $\dz \sim |\dphi|^{2b}$, and the bound \eqref{Domega} above $\omega^*$. These scalings persist until the anharmonic regime $T \sim |\dphi|^2$. Therefore we find simply
\eq{
\omega^* \sim \sqrt{T} |\dphi|^{-1/2} \qquad & (HS)
}
and 
\eq{ \label{msd2}
\langle \delta R^2 \rangle \gtrsim |\dphi|^{\kappa} \qquad & (HS)
} 
The results \eqref{msd1} and \eqref{msd2} display a striking asymmetry: the nontrivial temperature scale $T^*$, although it can be defined both above and below jamming, plays no role in the latter case. Physically, this is because when $\dphi >0$, there must be a temperature scale where the softest modes transition from `anomalous' to `sparse,' whereas when $\dphi <0$, the softest modes are always `sparse'. 

\section{Effective Medium} The scaling predictions of the previous section can be derived and extended with effective medium theory (EMT), a { mean-field } approximation to self-consistently treat disorder  \cite{Garboczi85,Webman81,Wyart10a,Mao10,DeGiuli14,Sheinman12}. Although EMT is formulated at $T=0$, an effective potential facilitates its use at any $T$. { An arbitrary distribution of disorder is allowed: here we assume that forces are distributed as Eq.\eqref{ex}, as in the variational bounds above. } We follow the same procedure as our previous works \cite{DeGiuli14,DeGiuli14b}, discussing all details in Appendix B.

In addition to reproducing the previous results, with values for prefactors, EMT also gives the behavior of the  shear modulus and density of states at all $\omega \ll 1$.  Here we focus on the results for $d=3$ at marginal stability, and only work to leading order in $\dz$. By numerical solution of the EMT equations at marginal stability, we obtain $\dz(T,\phi)$, plotted in Fig. \ref{fig:dz}. The asymmetry between $\dphi > 0$ and $\dphi <0$ is clearly visible. A curious feature is that at fixed $T$, $\dz$ has a minimum at a finite $\dphi$.  

The EMT prediction for $D(\omega)$ is plotted in Fig. \ref{fig:domega}a. Above $\omega^*$, it behaves as described in the foregoing text. Below $\omega^*$, we find $D(\omega) \sim \omega^2$; we note that our previous work \cite{DeGiuli14b} erroneously predicted $D(\omega) \sim \omega^{2+a}$ in this regime: an incorrect approximation was used to obtain scaling behavior below $\omega^*$, affecting only this prediction. 


EMT predicts that the zero-frequency shear modulus $\mu \propto -\dz + \sqrt{\dz^2 + C p}$, for a constant $C$. It can be checked that in all cases $p \lesssim \dz^2$, so that
\eq{ \label{mu1}
\mu \sim p/\dz,
}
consistent with arguments presented earlier in the $T \to 0$ limit \cite{Wyart08,Ellenbroek09,Ellenbroek06,DeGiuli14b}, and with the replica theory \cite{Yoshino14}. This implies
\eq{ \label{mu2}
\mu \sim \begin{cases} \dphi^{\frac{1}{2}} \qquad & (S0)\\
\dphi^{\kappa} T^{1-\kappa} \qquad & (SE) \\
T |\dphi|^{-\kappa} & (HS) \\
T^{1-\frac{\kappa}{2}}  & (AH),
\end{cases}
} 
plotted in Fig.\ref{fig:domega}b. In all cases we find
\eq{
\mu \langle \delta R^2 \rangle \sim T.
}
We note that this shear modulus only describes relaxation within a metastable state. 

Finally, EMT predicts the spatial correlation of mode displacements, $\ell_s(\omega)$. For $\omega \gtrsim \omega^*$, we find $\ell_s(\omega) \sim 1/\sqrt{\omega D(\omega)}$, leading to
\eq{ \label{ells}
\ell_s(\omega^*) \sim \begin{cases} \dphi^{-\frac{1}{4}} \qquad & (S0) \\
(T/\dphi)^{\frac{a-1}{4}} \qquad & (SE) \\
|\dphi|^{\frac{a-1}{4}} & (HS) \\
T^{\frac{a-1}{8}}  & (AH).
\end{cases}
} 
In principle, $\ell_s$ can be measured by static response, as shown previously for $T=0$ soft solids \cite{Lerner13a}.

\begin{figure}[t!] 
\includegraphics[width=0.5\textwidth,clip]{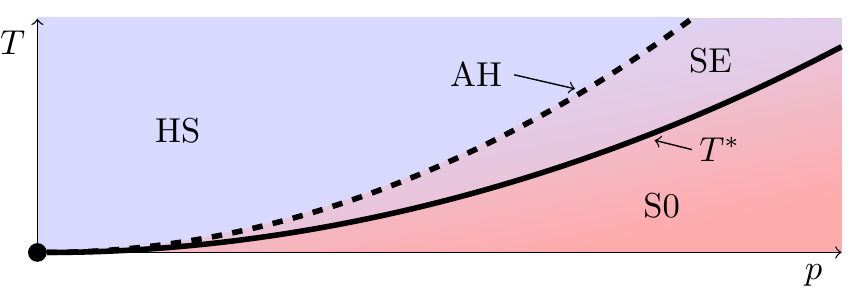}
%
%
\caption{\label{fig:phasediagram2} Phase diagram of Fig. \ref{fig:phasediagram} in $(p,T)$ space. Lines and colors are as in Fig. \ref{fig:phasediagram}. In these parameters, the regime AH folds into the curve $p \sim \sqrt{T}$. Note that the lines are schematic. }
\end{figure}

\section{Comparison with Simulations \& Experiments} Packings of soft harmonic spheres at finite temperature have been studied numerically by several groups \cite{Schreck11,Otsuki12,Henkes12,Ikeda12,Wang13,Ikeda13,Olsson13,Bertrand14}. 
A major goal of these works, particularly Refs. \onlinecite{Schreck11,Henkes12,Ikeda13}, was to show that vibrational properties become strongly anharmonic as packings are heated. As evidence, it was observed in Refs. \onlinecite{Schreck11,Henkes12,Ikeda13} that as a jammed solid at $\dphi > 0$ is heated, the plateau in $D(\omega)$ develops a negative slope at low frequency. We have shown that this phenomenon can be understood within a harmonic effective theory, and simply reflects the increasing importance of weak contacts (and therefore `sparse' modes) as the effective potential develops a small-force thermal tail. 

In Refs. \onlinecite{Ikeda13,Wang13,Bertrand14}, the three simple regimes HS, S, AH were observed, with the scaling $T \sim \dphi^2$ predicted here \footnote[4]{In Ref. \onlinecite{Bertrand14}, a dense liquid regime is also observed at larger $T$. Here our effective potential breaks down, as the system is no longer trapped in a metastable state.}. However, neither of these works found the subtle distinction between S0 and SE regimes. Because the temperature scale $T^* \sim |\dphi|^{2.20}$ (using $\theta_e=0.42$) is close to $\dphi^2$, precise measurements of $D(\omega)$, $\langle \delta R^2 \rangle$, or $\mu$ are needed to test our predictions. However our non-trivial prediction on the shape of the density of states (which can present a plateau or a power-law decay depending on the control parameters and the frequency) appears to be confirmed numerically, see e.g. Fig.2 in Ref.\onlinecite{Ikeda13} and Fig.3 in  Ref. \onlinecite{Wang13}.

{ Several colloidal experiments have aimed at measuring finite-temperature vibrational properties near jamming \cite{Zhang09,Ghosh10,Chen10}, although these observations may not be close enough to the critical point to obtain accurate exponents\cite{Jacquin10,Ikeda13}. However, recent experiments on granular media\cite{Coulais12,Coulais14} may probe this regime, albeit in the presence of friction. In Refs. \onlinecite{Coulais12,Coulais14}, heterogeneity associated with dynamics of contacts was used to infer Widom lines emerging from a common point at $T=0, \phi=\phi_c$, consistent with the results of this work. A more detailed comparison could be possible by computing dynamical susceptibility, along the lines of Ref. \onlinecite{Ikeda13}. 
}

\section{Conclusion}

We have proposed a  description of the vibrational properties of thermal soft spheres near the jamming transition, based on a real-space description of   soft vibrational modes. These modes are of two types -- sparse and anomalous-- depending on packing fraction, temperature and on the frequency considered. Our work shows that even when anharmonicities are extremely strong, they can be tamed by time-averaging,  allowing one to define an effective potential from which static linear response and thermal fluctuations can be computed. Ultimately all our scaling predictions can be expressed in terms of one exponent $\theta_e$ characterizing forces in jammed packings, which can be extracted numerically. 

$\theta_e$ is also very well estimated by replica calculations in infinite dimensions \cite{Charbonneau14}. 
In the hard sphere case ($\phi<\phi_c$, $T\rightarrow 0$) such calculations  lead to identical results for the  mean-square displacement \cite{Charbonneau14} and the elastic modulus \cite{Yoshino12,Yoshino14}. However this approach currently does not fix the exponents for soft spheres \cite{Yoshino14}, nor does it access the shape of the vibrational spectrum nor the length scale characterizing normal modes. It would be very interesting to extend the replica method to these questions, enabling a full comparison with the present results. 

Although we have focused on the proximity of the jamming transition, we expect our description based on an effective potential to be qualitatively appropriate away from it. For example, if the packing fraction is decreased (or the temperature increased), an elastic instability is expected to occur \cite{Brito09,DeGiuli14,Lerner14} near the glass transition where configurations become unstable. Denoting $\phi_0$ such a packing fraction, effective medium\cite{DeGiuli14} predicts that the elastic modulus jumps and displays a square-root singularity at that point $\mu\approx \mu_0+C\sqrt{\phi-\phi_0}$, as also found in Mode Coupling Theory \cite{Gotze85,Szamel11}, replica theory \cite{Yoshino12,Yoshino14}, and the random-first-order-transition scenario \cite{Kirkpatrick89}. Furthermore, numerics support that an elastic length scale diverges in two dimensions as\cite{Lerner14} $l_c\sim 1/(\phi-\phi_0)^{1/4}$ and therefore decreases as the system enters  the glass phase.  We expect these behaviors to be smoothed to some extent, since in the vicinity of the glass transition the separation of time scales required to define an effective potential disappears. Direct extraction of elastic length scales as a function of packing fraction or temperature would be very informative to test this prediction. 

Finally, one may wonder what happens to the soft sphere phase diagram when the system is driven by an imposed shear. For $T=0$ and $\phi<\phi_c$ it has been shown that the dynamics of over-damped hard spheres is governed by an operator whose low-energy modes are of the anomalous  type \cite{Lerner12}, so that velocity correlations decay primarily on a length scale  $\ell_c\sim 1/\sqrt{z_c-z}$ independent of $\theta_e$ \cite{During13,During14}.  It would be interesting to see if the notion of effective potential could be used to extend such considerations to finite temperature, a path recently proposed in Ref. \onlinecite{Trulsson14}.

 {\bf Acknowledgments}: It is a pleasure to thank G. Biroli, G. D\"uring, A. Grosberg, J. Lin, L. Yan, H. Yoshino, and F. Zamponi for discussions related to this work, and we especially thank the authors of Ref. \onlinecite{Henkes12} for sharing their data. MW acknowledges support from NSF CBET Grant 1236378, NSF DMR Grant 1105387, and MRSEC Program of the NSF DMR-0820341 for partial funding. 

\vfill
\pagebreak

\appendix

In these Appendices we (A) derive the effective potential, and (B) describe the effective medium theory.

\renewcommand{\theequation}{A.\arabic{equation}}
\setcounter{equation}{0}

\section{Effective Potential}

Here we derive, in more detail, the effective potential used in the main text. We consider a system of $N$ soft spheres at constant pressure $p$ that is fluctuating around an isostatic metastable state. The latter has a Gibbs partition function
\eq{ \label{Z0}
Z(p,\beta)= \int D\delta \rb \; e^{-\beta p V} e^{-\beta U},
} 
where $U$ is elastic energy, $V$ is volume, $p$ is pressure, $\int D\delta \rb \equiv \prod_i \int d\delta \rb_i$, and $\beta=1/(k_B T)$. The displacements are with respect to a configuration in which all particles are just touching, at volume $V_c$.

Metastability of the state is interpreted to mean that although the state may have a finite lifetime, the time-averaged positions of the particles are in force balance. Then there exist contact forces $\{ f_\alpha \}$ balancing the external forces $\{\vec{F}_i \}$:
\eq{
\vec{F}_i = \sum_{\alpha \in i} f_\alpha \langle\nb_\alpha\rangle,
}
for each $i$, where $\langle\nb_\alpha\rangle = \langle \rb_j - \rb_i \rangle/|\langle \rb_j - \rb_i \rangle|$ is the time-averaged contact vector. Summing over all particles and contracting with an arbitrary displacement field, we get an expression for the work done in the displacement field:
\eq{
W = \sum_i \vec{F}_i \cdot \delta \rb_i & = - \sum_\alpha f_\alpha \langle\nb_\alpha\rangle \cdot (\delta \rb_j - \delta \rb_i) \notag \\
& = - \sum_\alpha f_\alpha (h_\alpha + \OO(h_\alpha^2)), \label{virtwork}
} 
the virtual work principle \cite{Roux00}. Here we have defined the gap $h_\alpha$ between particles, $h_\alpha = \nb_\alpha \cdot (\delta \rb_j - \delta \rb_i)$ (with $h_{\alpha}<0$ for overlap). The error in Eq.\eqref{virtwork} comes from the deviation of time-averaged contact vectors from instantaneous ones; this is small if particle displacements are also small, $|\delta \rb_i | \ll 1$. Because of isostaticity, each force $f_\alpha$ in this expression can be written $f_\alpha = p {\hat f_\alpha}$, where ${\hat f_\alpha}$ is fixed on the mesoscopic time $\tau$ and depends on the time-averaged positions of the particles. The pressure simply sets the force scale. 
  
Isostaticity implies that the number of degrees of freedom in the particle displacements $\{\delta \rb_i \}$, $dN$, is precisely equal to the number of contacts, $N_C=zN/2$. Thus Eq.\eqref{Z0}, originally an integral over the displacements $\{\delta \rb_i\}$, can be written as an integral over the gaps $\{h_{\alpha}\}$ between contacting particles, i.e., 
\eq{
Z(p,\beta)= \int Dh \; J \; e^{-\beta p V} e^{-\beta U}
} 
where the Jacobian $J = |$constant$ + \OO(h)|$: to leading order, $J$ depends only on the time-averaged positions of the particles, with a small correction due to contact vector rotation. Since compressing or dilating the system from volume $V_c$ to $V$ requires work $W=-p(V-V_c)$, we can substitute Eq.\eqref{virtwork} into $Z$ to obtain
\eq{
Z(p,\beta)= J e^{-\beta pV_c} \prod_\alpha \int dh_\alpha \; e^{-\beta f_\alpha h_\alpha} e^{-\beta U(h_\alpha)},
} 
where we assume that $U$ is a pair potential, thus depending only on $h_\alpha$. Strictly, each gap $h_\alpha$ should be bounded from above by some nontrivial function of the positions of the rest of the particles. However, since we are only interested in small displacements, we ignore this fact and let the upper limit on each $h_\alpha \to \infty$: large gaps are strongly suppressed by the Boltzmann factor. ( {\it A posteriori}, one can check the smallness of temperature needed for gaps to remain $\ll 1$.) Finally, we are led to a single-gap partition function \cite{Wyart05b,Brito06,Brito09}
\eq{ \label{Z2}
Z(\beta,f) = e^{-\beta G} = \int dh \; e^{-\beta U(h)} e^{-\beta f h},
}
where the force $f$ fixes the time-averaged gap $\lh$ through $\lh = \p G/\p f$. The effective potential $V_{\eff}$ should be a function of the time-averaged gap $\lh$; we define it by Legendre transform $V_{\eff}(\lh) = G(f)-f \lh$, from which it follows that $f = -\p V_{\eff}(\lh)/\p \lh$, as desired. We consider a finite-range harmonic potential $U(h)=\half\epsilon |h/\sigma|^2$ when $-\half \sigma < h < 0$, and 0 otherwise. We now take units in which $\epsilon=\sigma=k_B=1$. 

Eq.\eqref{Z2} and $f = -\p V_{\eff}(\lh)/\p \lh$ define implicitly the effective potential. For our choice of $U(h)$, we obtain an error function:
\eq{
Z & = \int_0^\infty dh \; e^{-\beta f h} + \int_{-1/2}^0 dh \; e^{-\beta f h} e^{-\half \beta |h|^{2}} \\
& = \frac{1}{\beta f} + \frac{1}{\sqrt{\beta}} e^{\half \beta f^2} \int_{\sqrt{\beta} (f-\half)}^{\sqrt{\beta} f} dz \; e^{-\half z^2}
}
We are interested in $\beta \to \infty$ so $\sqrt{\beta} (f-\half) \approx -\infty$. Then the equation defining the effective potential is
\eq{ \label{AZ}
\beta \lh & = -\frac{\p \ln Z}{\p f} \\
& = \frac{-1}{Z} \left[ -\frac{1}{\beta f^2} + 1 + \beta^{1/2} f e^{\half \beta f^2} \int_{-\infty}^{\sqrt{\beta} f} dz \; e^{-\half z^2} \right] \notag
}
We see that the behavior of Eq.\eqref{AZ} depends on the scaling variable $\sqrt{\beta}f$. When $\beta f^2 \gg 1$, the contact term dominates, while when $\beta f^2 \ll 1$, the entropic term dominates. The integral is always $\OO(1)$, so we approximate it by $\sqrt{\pi}/2$. 

{\it Entropic regime: } Let $\beta f^2 \ll 1$. Then $Z \approx 1/(\beta f)$ and 
\eq{
\lh \approx \frac{1}{\beta f}
}
This recovers the hard-sphere effective potential derived in Ref.\onlinecite{Brito09}. It holds when $f \ll 1/\sqrt{\beta}$, or $\lh \gg 1/\sqrt{\beta}$.

{\it Contact regime: } Let $\beta f^2 \gg 1$. Then $Z \approx \sqrt{\pi/\beta}/2 \; e^{\half \beta f^2}$ and 
\eq{
\lh \approx -f
}
This recovers the original harmonic potential. It holds when $f \gg 1/\sqrt{\beta}$, or $\lh \ll -1/\sqrt{\beta}$.

Therefore there is a characteristic scale $1/\sqrt{\beta}=\sqrt{T}$ both for gaps and overlaps. We cannot analytically solve the equations in the nontrivial regime $-\sqrt{T} < \lh < \sqrt{T}$ but there is a smooth cross-over (for example, there is no $f$ with $d\lh/df=0$, since $d\lh/df \propto -\langle (h-\lh)^2 \rangle < 0$). 

In the main text, we want to use $V_{eff}$ to obtain scaling behavior of vibrational properties; we can do so by replacing the cumbersome implicit expression  with a simple form that incorporates the correct limiting behavior. The simplest is the one satisfying $\lh = -f + T/f$, which leads to an effective force law
\eq{ \label{Afeff}
f(\lh) = -\half \lh + \half \sqrt{\lh^2+4T},
}
from which we can compute the effective potential $V_{\eff}(\lh) = -\half\lh f(\lh) - T \log[\lh+\sqrt{\lh^2+4T}]$ and stiffness
\eq{ \label{Akeff}
k(\lh) \equiv & -\frac{df}{d\lh} = \frac{1}{2} - \frac{\lh}{2\sqrt{\lh^2+4T}}
}
These relations are valid for $T \ll f^2$ and $T \gg f^2$, but perhaps not close enough to $T \sim f^2$. 

\renewcommand{\theequation}{B.\arabic{equation}}
\setcounter{equation}{0}

\section{Effective Medium Theory}

Effective medium theory (EMT) is a self-consistent approximation scheme to treat disorder. Here we follow our previous works \cite{DeGiuli14,DeGiuli14b}, the only difference with respect to Ref. \onlinecite{DeGiuli14b} being the effect of softness on the effective potential Eq.\eqref{Afeff} (in that work we had $f(\lh) = T/\lh$). Accordingly, we only sketch the derivation of our results. 

{\it Equations:}
We assume that contact forces have a distribution
\eq{
\PP(f) =C_f f^{\theta} e^{-f/\fbar},
}
and that forces and bond stiffnesses follow the effective potential Eq.\eqref{Afeff}, Eq.\eqref{Akeff}. {  We fix the bond length to one, since incorporating the dependence of bond length with force leads to subdominant corrections near jamming.} A random elastic network of coordination $z$ is constructed by random dilution of a regular lattice of coordination $z_0$ down to $z$. The stiffness in contact $\alpha$, $k_\alpha$, and the force in the contact, $f_\alpha$ are random variables distributed according to
\eq{
\PP_{EMT}(k_\alpha) & = (1-P) \delta(k_\alpha) + P \; \PP(k_\alpha) \\ 
\PP_{EMT}(f_\alpha) & = (1-P) \delta(f_\alpha) + P \; \PP(f_\alpha) 
}
where $P=z/z_0$. The bond stiffness distribution is $\PP(k) = df/dk \; \PP(f)$. In EMT, the elastic behavior of a random material is modelled by a regular lattice with effective frequency-dependent stiffnesses; here we will have a longitudinal stiffness, $\kpa$, and a transverse stiffness $-\kpe$. Writing $\overline{\; \cdot \; }$ for disorder average, the EMT equations are
\eq{
0 = \overline{ \frac{\kpa-k_\alpha}{1-(\kpa-k_\alpha) G^\parallel} } = \overline{ \frac{\kpe-f_\alpha}{1+(\kpe-f_\alpha) G^\perp} },
}
where $\Gpa$ and $\Gpe$ are related to the Green's function $\Gb(\omega) = \big(\curlyM - m\omega^2\big)^{-1}$ by
\eq{
G^\parallel & = \nb_\alpha \cdot \langle \alpha | \Gb | \alpha \rangle \cdot \nb_\alpha \\
G^\perp & = \frac{1}{d-1} \left[ \tr( \langle \alpha | \Gb | \alpha \rangle) - G^\parallel \right],
}
with $\langle \alpha | \equiv \langle i | - \langle j|$. Using Eq.\eqref{Afeff} one obtains the EMT equations
\eqs{
0 & = \frac{(1-P) \kpa}{1-\kpa G^\parallel} + P \left[ \frac{-1}{\Gpa} + \frac{1}{\Gpa{}^2(1+c)} \right. \notag \\
& \qquad \times \left. \left(1 + \frac{C_f T \fbar^{\theta-1}}{1+c} \int df\; \frac{f^\theta e^{-f}}{\tilde{c}T\fbar^{-2}+f^2} \right) \right] \\
0 & = \frac{(1-P) \kpe }{1+\kpe G^\perp} + P \left[\frac{1}{G^\perp} - \frac{C_f \fbar^{\theta}}{{G^\perp}^2} \int_0^\infty df \; \frac{f^\theta e^{-f}}{c_2 - f}  \right],
}
where $c=(1-\kpa \Gpa)/\Gpa$, $\tilde{c}=c/(1+c)$, and $c_2 = (1+\kpe \Gpe)/(\fbar \Gpe)$. Near the Maxwell point we expect $|c|\ll 1$ and $c_2 \gg 1$ (which can be checked {\it a posteriori}), leading to 
\eq{ \label{Aeqns}
0 & = \kpa \Gpa - P + \frac{(1-\kpa \Gpa) P}{\Gpa} \notag \\
& \qquad \times \left[ 1 + C_1 T \fbar^{-1-\theta} \left( \frac{T(1-\kpa \Gpa)}{\Gpa} \right)^\alpha + \ldots \right] \\
0 & = \kpe \Gpe - P \frac{\fbar \Gpe (\theta + 1)}{1 + \kpe \Gpe} + \ldots \label{Aeqns2}
}
with $\alpha = (\theta-1)/2$, $C_1 = \pi/(2\Gamma_{1+\theta} \sin|\pi \alpha|)$, $\Gamma_t=\int_0^\infty x^{t-1} e^{-x} dx$. 

For the Green's function we consider
\eq{ 
\Gb(\rb,\omega) = \delb \int_{BZ} \frac{d^d q}{\rho (2\pi)^d} \frac{e^{i \qb \cdot \rb}}{(\kpa- \kpee) q^2 - m \omega^2},
}
where $BZ = \{ \qb: \; |\qb| < \Lambda  \}$ is an approximate first Brillouin zone, $\delb$ is the identity tensor, $\rho=N/V$ is the number density, and $\kpee = (d-1)\kpe$. Isotropy of $\Gb$ implies an identity
\eq{ \label{Aeqns3}
G^\parallel = G^\perp & = \frac{2d}{z_0} \frac{1}{\kpa-\kpee} \left( 1 + \frac{m \omega^2}{d} \tr(\Gb(0,\omega)) \right).
}
Assuming $\omega \ll \sqrt{\kpa/m}$ and $d \geq 3$, it can be checked that
\eq{ \label{Aeqns4}
\frac{1}{d} \mbox{tr}[\Gb(0,\omega)] = \frac{A_1}{\kpa-\kpee} + \ldots
}
with
\eq{
A_1 = \frac{2 \pi^{d/2}}{\rho \Gamma_{d/2} (2\pi)^d} \frac{\Lambda^{d-2}}{d-2}
}
Eqns.\eqref{Aeqns},\eqref{Aeqns2},\eqref{Aeqns3},\eqref{Aeqns4} are solved for $\dz \ll 1$ by supposing
\eq{
\kpa & \sim \dz^\xi, \quad \kpe \sim \dz^\eta, \\
\omega & \sim \dz^\zeta, \quad T \sim \dz^\nu, \quad \fbar \sim \dz^\gamma,
}
and balancing terms in the above equations. This necessarily leads to a description of the behavior near $T^*$, where both hard-sphere and soft-sphere physics is present at small frequency. We find
\eq{
\eta=\gamma=2\xi=2\zeta=2, \quad \nu = \frac{5+3\theta}{1+\theta},
}
recovering the scalings described in the main text. 

{\it Results:} The leading order equations for $\kpe, \kpa$ are
\eq{
\kpe & = P \fbar (1+\theta) \sim p
}
and
\eq{ \label{Amaineqn}
0 = z_c A_1 \omega^2 - \kpa \dz + \kpee z_c + a \kpa{}^2 + c_3 \kpa \left(\frac{T\kpa}{ \fbar^2}\right)^{\alpha+1}  
}
with $a=z_0-z_c$ and $c_3 = C_1 z_0 (1-z_c/z_0)^{\alpha+1} (z_0/z_c)^{\alpha}.$ This equation contains the leading-order terms (in $\dz$) for both the harmonic $T=0$ theory from Ref. \onlinecite{DeGiuli14} and the hard-sphere theory from Ref. \onlinecite{DeGiuli14b}. When all terms are important, it describes behavior near $T^*$. Other regimes are obtained when one or several terms becomes small. We cannot analytically solve Eq.\eqref{Amaineqn}, but we can determine some of its key properties. 

For example, we expect $\kpa(\omega)$ to have an onset frequency $\omega_0$ where the density of states rises from zero. At this point we must have $|d\kpa/d\omega|=\infty$, implying
\eq{ \label{Aom01}
0 = \dz - 2a \kpa_0 - (\alpha+2)c_3 \left(\frac{T\kpa_0}{ \fbar^2}\right)^{\alpha+1},
}
and
\eq{ \label{Aom02}
0 = \gamma_2 \kpa_0{}^2 + \gamma_1 \dz \kpa_0 - z_c (A_1 \omega_0^2 + \kpee),
}
where $k_0=\kpa(\omega_0)$, with $\gamma_1 = (1+\theta)/(3+\theta)$, $\gamma_2 = a(1-\theta)/(3+\theta)$. $k_0$ and $\omega_0$ can be found by numerical solution of Eqs.\eqref{Aom01},\eqref{Aom02} for a prescribed value of $\dz$.

{\it Marginal stability:} The condition of marginal stability corresponds to\cite{DeGiuli14} $\omega_0=0$. Therefore $\dz(p,T)$ at marginal stability can be found by solving Eqs.\eqref{Aom01},\eqref{Aom02} in this case, leading to Fig.\ref{fig:dz} in the main text. Since $k_0=\kpa(\omega_0)$, at marginal stability $k_0-\kpee = k_0 (1 + \OO(\dz))$ is precisely the shear modulus $\mu$. Hence
\eq{
\mu = \frac{\gamma_1}{2\gamma_2} \left[ -\dz + \sqrt{\dz^2 + \gamma_3 \fbar} \right],
}
as discussed in the main text, and plotted in Fig.\ref{fig:domega}b.

{\it Density of states: } The density of states 
\eq{ \label{ADom}
D(\omega) & =(2\omega/\pi)\mbox{Im[tr[}\Gb(0,\omega)]] \notag \\
& = \frac{2d A_1 \omega}{\pi} \mbox{Im}[1/\Delta k] + \ldots \notag \\
& = -\frac{2d A_1 \omega}{\pi} \frac{\mbox{Im}[\Delta k]}{|\Delta k|^2} + \ldots
}
with $\Delta k = \kpa - \kpee$. We determine it by numerically solving Eq.\eqref{Amaineqn} at each $\omega$. It is plotted in Fig.\ref{fig:domega}a. 

{\it Scattering Length: } Finally, we can extract the asymptotic behavior of the Green's function for large $r$. To leading order, $\log(\Gb(r,\omega))\sim -r/\ell_s(\omega) + i \omega r/\nu(\omega)$ where $\ell_s(\omega)=-\omega^{-1} |\Delta k|/\mbox{Im}[\sqrt{\Delta k}]$ and $\nu(\omega)=|\Delta k|/\mbox{Re}[\sqrt{\Delta k}]$ are, respectively, the scattering length and sound velocity at frequency $\omega$.

To extract $\ell_s$ at the characteristic frequency $\omega^*$, we use our knowledge of the shape of $\Delta k \sim \kpa$ known from our previous works \cite{DeGiuli14,DeGiuli14b}, which is confirmed by numerical solutions to Eq.\eqref{Amaineqn}: near $\omega^*$, we have Re$[\kpa] \sim $ constant $\sim \mu$, while $-$Im[$\kpa] \gtrsim $Re[$\kpa]$. This implies that $|\Delta k| \sim $-Im[$\kpa]$. Using $2 $Im${}^2\sqrt{\Delta k}= |\Delta k| - $Re$[\Delta k] \approx |\Delta k|$ along with the definitions of $\ell_s$, and Eq.\eqref{ADom}, we find
\eq{
\ell_s(\omega^*) & = -\frac{|\Delta k|}{\omega^* \mbox{Im}[\sqrt{\Delta k}]} \notag \\
& \sim -\frac{|\Delta k|^{1/2}}{\omega^*} \notag \\ 
& \sim [ D(\omega^*) \omega^* ]^{-1/2},
}
from which we can determine the scaling properties of $\ell_s(\omega^*)$.

\bibliographystyle{apsrev4-1}
\bibliography{
../bib/Wyartbibnew}{}
\end{document}